# Bolometric Bond Albedo and Thermal Inertia Maps of Mimas

C.J.A. Howett[1], J.R. Spencer[1] and T. Nordheim[2]

1 - Southwest Research Institute, Colorado, USA.

2 – Jet Propulsion Laboratory, California, USA

**Corresponding Author and their Contact Details:**

C.J.A. Howett

Email: howett@boulder.swri.edu

Telephone Number: +1 720 240 0120

Fax Number: +1 303-546-9687

Address:

1050 Walnut Street, Suite 300

Boulder, Colorado

80302

USA

**Abstract**

In 2011 a thermally anomalous region was discovered on Mimas, Saturn's innermost major icy satellite (Howett et al., 2011). The anomalous region is a lens-like shape located at low latitudes on Mimas' leading hemisphere. It manifests as a region with warmer nighttime temperatures, and cooler daytime ones than its surroundings. The thermally anomalous region is spatially correlated with a darkening in Mimas' IR/UV surface color (Schenk et al. 2011) and the region preferentially bombarded by high-energy electrons (Paranicas et al., 2012, 2014; Nordheim et al., 2017).

We use data from Cassini's Composite Infrared Spectrometer (CIRS) to map Mimas' surface temperatures and its thermophysical properties. This provides a dramatic improvement on the work in Howett et al. (2011), where the values were determined at only two regions on Mimas (one inside, and another outside of the anomalous region). We use all spatially-resolved scans made by CIRS' focal plane 3 (FP3, 600 to 1100 cm$^{-1}$) of Mimas' surface, which are largely daytime observations but do include one nighttime one. The resulting temperature maps confirm the presence and location of Mimas' previously discovered thermally anomalous region. No other thermally anomalous regions were discovered, although we note that the surface coverage is incomplete on Mimas' leading and anti-Saturn hemisphere. The thermal inertia map confirms that the anomalous region has a notably higher thermal inertia than its surroundings: 98±42 J m$^{-2}$ K$^{-1}$ s$^{-1/2}$ inside of the anomaly, compared to 34±32 J m$^{-2}$ K$^{-1}$ s$^{-1/2}$ outside. The albedo inside and outside of the anomalous region agrees within their uncertainty: 0.45±0.08 inside compared to 0.41±0.07 outside the anomaly. Interestingly the albedo appears brighter inside the anomaly region, which may not be surprising given this region does appear brighter at some UV wavelengths (0.338 μm, see Schenk et al., 2011). However, this result should be treated with caution because, as previously stated, statistically the albedo of these two regions is the same when their uncertainties are considered. These thermal inertia and albedo values determined here are consistent with those found by Howett et al. (2011), who determined the thermal inertia inside the anomaly to be 66±23 J m$^{-2}$ K$^{-1}$ s$^{-1/2}$ and <16 J m$^{-2}$ K$^{-1}$ s$^{-1/2}$ outside, with albedos that varied from 0.49 to 0.70.

**1 Introduction**

Mimas is the smallest of Saturn's major icy satellites (with a mean radius of 198 km), and the innermost (semi-major axis of 185,539 km). In visible light its surface appears heavily cratered and bland. So it was somewhat a surprise when in 2011 analysis of data obtained by Cassini's Composite Infrared Spectrometer (CIRS) showed that there was a thermal anomaly at low latitudes on Mimas' leading hemisphere (Howett et al., 2011; Figure 1). The temperature in the anomalous region was shown to be warmer at nighttime and cooler during the day than its surroundings by ~15 K. The cause of the anomaly was shown to be due to the surface having a higher thermal inertia (66±23 J m$^{-2}$ K$^{-1}$ s$^{-1/2}$, these units are henceforth referred to as MKS) compared to its surroundings (<16 MKS). Thermal inertia is defined as being $\sqrt{k\rho c}$, or $\sqrt{k\rho_0(1-p)c}$ where $k$ is the thermal conductivity, $\rho$ is the density, $c$ is the specific heat, $\rho_0$ is the zero porosity density and $p$ is the porosity.

Mimas' thermal anomaly is spatially correlated with a color anomaly (c.f. Schenk et al., 2011), which is observed as a darkening in the IR (0.930 μm) /UV (0.338 μm) color ratio (Figure 2). Both the color and thermal anomaly are also spatially correlated with the region preferentially bombarded by high-energy electrons (c.f. Paranicas et al., 2012, 2014; Nordheim et al., 2017). The leading hypothesis to explain both the color and thermal anomalies is that these high-energy electrons modify Mimas' surface, mobilizing water ice grains to increase grain-to-grain contact (in effect gluing the grains together, increasing their effective thermal conductivity and hence effective thermal inertia) and adding scattering centers which enhance short-wavelength reflectivity (c.f. Schenk et al., 2011; Howett et al. 2011; Schaible et al., 2016).

In the original thermal anomaly discovery paper (Howett et al., 2011) the thermal inertia and albedo were determined at just two locations (one inside and the other outside of the thermally anomalous region). Here we expand on that work, to map Mimas' thermophysical properties across its surface.

**2 Data**

All of the observations used in this analysis were made by the Composite Infrared Spectrometer (CIRS) onboard the Cassini spacecraft (Flasar *et al*., 2004). CIRS is a Fourier Transform Spectrometer (FTS) that has two interferometers that share a single scan mechanism and telescope. Low wavenumbers 10 to 600 cm$^{-1}$ (16.7 to 1000 μm wavelengths) are detected by two thermopile detectors known as focal plane 1 (FP1), which has a spatial resolution of 3.9 mrad. Higher wavenumbers 600 to 1100 cm$^{-1}$ (9.1 to 16.7 μm) and 1100 to 1400 cm$^{-1}$ (7.1 to 9.1 μm) are detected by focal planes 3 and 4 respectively (known as FP3 and FP4). These two focal planes each have a row of ten square detectors, each of which has a 0.273 by 0.273 mrad field of view.

The wavelength range of FP1 makes it sensitive to temperatures above ~25 K, meaning that it is sensitive to both the day and nighttime surface temperatures of the icy Saturnian satellites. FP3 is only sensitive to temperatures above ~65 K, meaning that it is only sensitive to daytime surface temperatures and some anomalous regions (e.g. the warm nighttime temperatures in Mimas' thermally anomalous region, and Enceladus warm active regions). The sensitivity of FP4 to day and nighttime surface temperatures of the icy Saturnian satellites is too low to be of use, with the exception of Iapetus and the warm active regions on Enceladus. Since FP3 has a higher spatial resolution than FP1, and it is sensitive to daytime temperatures and nighttime temperatures inside Mimas' thermally anomalous region, we focus on data taken by this focal plane in this work.

CIRS observed Mimas over a total of 21 Cassini orbits (Revs). We use observations from three of these observations, which were chosen as they provide the highest spatial resolution coverage of Mimas' anomalous region for FP1 (nighttime) and FP3 (daytime). An overview of the observations analyzed is given in Table 1. As the table shows four separate scans being analyzed from the first observation sequence (Rev 12). Most of the observations are taken at low phase (i.e. daytime observations), with the exception being

Rev 139, which was taken at a phase angle of 114º. The data was restricted to those with an emission angle less than 60º to minimize the effect of roughness. Table 2 provides a brief outline of CIRS' observations of Mimas that were not used in this study, which are included to help future researchers in their CIRS selection for Mimas.

3 Data Analysis and Results

For each sequence (i.e. each line in Table 1) the CIRS observations were binned into 10° by 10° longitude and latitude bins. In each bin the mean spectrum is found, and then the closest blackbody fit to that spectrum is assumed as the bin's temperature. The blackbody fit was determined by fitting a blackbody curve to the observed emission using the downhill simplex method (c.f. Nelder and Mead, 1965) in IDL's amoeba algorithm, on the assumption that the surface emits as a blackbody.

The noise on the derived surfaces temperatures is derived using a two-step Monte Carlo technique: first a synthetic noise with a comparable magnitude to the observed noise is created and added to the previously determined best fitting blackbody curve. Then this spectrum is fitted by a blackbody emission spectrum. This process is repeated numerous times, and the temperature error estimate is given by the standard deviation of the temperatures whose blackbody emission spectra are best able to fit the created spectra. This process is repeated for all observations listed in Table 1, and the results are shown in Figure 3. As the Figure shows the errors on some temperature fits are quite large (as high as 30 K on temperatures ~75 K). To prevent these large errors from propagating into the fitted thermal inertia and albedo values we only use derived temperatures that have a low error value (arbitrarily set at better than ±5 K). These larger errors are because of FP3's decreased sensitivity to low surface temperatures. As already stated, FP3 is only sensitive to temperatures warmer than ~65 K. As the surface temperature approaches such cold values the CIRS spectra get increasingly noisy and thus the error associated with temperatures derived from them increases.

We use the 1-D thermal model described fully in Spencer (1989) known as *thermprojrs*. The model calculates, in one-dimension, the heat flow conducted to and away from the surface to determine the temperature as a function of depth and time of day. The upper boundary is set so that the thermal radiation and incident solar radiation are balanced with the heat conducted to and from the surface and the change in the heat content of the surface layer. The lower boundary is set to a depth at which there is negligible temperature change with the diurnal temperature cycle. The model was run for a range of albedo and thermal inertia values, and for the specific geometry of each sequence (i.e. accounting for the rotation speed of the target, its heliocentric distance, latitude, local time and subsolar latitude). Based on the results of Howett et al. (2010, 2011) bolometric Bond albedos were sampled between 0.30 and 0.74 at increments of 0.02, and thermal inertias between 1 and 200 MKS at 2 MKS intervals. We assume a emissivity of unity, which has been shown to be reasonable for observing icy satellite surfaces at CIRS wavelengths (e.g. Carvano et al., 2007; Howett et al, 2016).

The predicted surface temperatures for each observation are then compared to those derived from binned CIRS observations. In some nighttime observations the signal of CIRS' FP3 detector is so low that a temperature cannot be determined. In these cases we assume an upper-limit of 65 K for the surface's temperature. Since a constraint of surface emission is set by temperature we conduct all of the analysis in "temperature-space" (i.e. not radiance). In each latitude/longitude bin an albedo and thermal inertia combination is considered good if it can reproduce the observed temperatures for two observation sequences (to within the observational error). For each bin the mean and standard deviation of both parameters are then calculated from the good albedo and thermal inertia combinations found for that bin. These results are then mapped, as shown in Figure 4.

An example of this fitting procedure is given in Figure 5 for two bin locations, one inside the anomalous region (130° to 140° W, 0° to 10° N) and another outside of it (200° to 210° W, 0° to 10° N). The figure shows the surface temperatures derived from CIRS observations (with their associated error, and the range in local time the bin spans) compared to model temperatures (and their associated albedo and thermal inertias) that are able to fit the observed temperatures. Finally the albedo and thermal inertia combinations that both observations are able to fit are also given.

Figure 5 clearly shows the importance of using surface temperatures at different local times to constrain a surface's albedo and thermal inertia. In the 120 to 130° W bin a morning and afternoon observation were used, compared to two mid-afternoon observations at 210° to 220° W. When different local times observed, the resulting thermal constraint and albedo is much better. For example for the bin (120 to 130° W, 0° to 10° N) the thermal inertia and albedo are 115±20 MKS, 0.52±0.01 respectively, compared to the bin (210° to 220°, 0° to 10° N) which has thermal inertia and albedo values of 118±46 MKS, 0.49±0.07. However, since only limited CIRS data is available it was not possible to use data from notably different local times in some bins. These differences account for the wide range in error bars on the derived thermal inertia and albedo. As Figure 4 shows, the mapped thermal inertias mostly have error bars less than ±40 MKS (but some reach as high as ±70 MKS), and albedos are all correct to better than ±0.15.

**4 Discussion**

The CIRS data we used (i.e. FP3) did not provide enough coverage of Mimas' trailing and Saturn-facing hemisphere to allow the thermal properties of these regions to be mapped. While the maps presented here are the most complete made to date there are still large areas of Mimas' surface that remain unmapped. Despite these gaps the "PacMan" anomaly on Mimas, first discovered by Howett et al. (2011) is clearly seen in both the temperature maps (Figure 3) and Mimas' thermal inertia map (Figure 4). No other thermally anomalous regions were observed.

Paranicas et al. (2014) used a combination of modeling and data from Cassini's Magnetosphere Imaging Instrument (MIMI) to predict patterns of electron energy deposition. If assume these results, and that the boundary of the thermally anomalous region coincides with where the energetic electron power deposited into the surface per unit area is $5.6 \times 10^4$ MeV cm$^2$ s$^{-1}$ (i.e. the dotted line in Figures 3, c.f. Howett et al. 2011) then the bins that are inside the anomalous region are shown in Figure 3. The mean thermal inertia of these bins is 98±42 MKS inside the anomaly, compared to 24±32 MKS outside (see below for further discussion of these values).

Diurnal temperatures probe the thermophysical properties over the thermal skin depth $\delta$, which is defined as $\delta = I/\rho c \sqrt{\omega}$ or where I is thermal inertia, $\rho$ is surface density, c is specific heat and $\omega$ is the angular velocity of rotation (i.e. spin). Assuming the specific heat is the same as water ice at 90 K (0.8 J K$^{-1}$ g$^{-1}$), and the density is the same as non-porous ice at 93 K (0.934 g cm$^{-3}$) and Mimas' rotation rate of 0.942 Earth days then the thermal skin depth inside and outside of the thermally anomalous region is 1.54 cm and 0.41 cm respectively. If we substitute the earlier expression for thermal inertia then skin depth can be shown to be inversely proportional to one minus the porosity:
de depth can be shown to be inversely proportional to one minus the porosity: $\delta = k\rho$ $\delta = \sqrt{k/\rho_0(1-p)c\omega}$. So as the porosity increases so to does the skin depth.

Due to the large uncertainties on the albedo the values outside and inside of the thermally anomalous region overlap (see Figure 7). The albedos inside and outside of the thermally anomalous region appear to follow the same trend, being brightest in equatorial regions and darker at higher latitudes. However, it appears in Figure 3 that the albedo inside of the anomalous region is brighter than outside it, with a boundary that appears to have the same shape as the thermal inertia change. This at first appears to be in contrast to previous studies, which have found Mimas to be brighter on its trailing hemisphere (e.g. Verbiscer and Veverka, 1992; Buratti et al., 1998). However, our surface coverage outside of the thermally anomalous region covers the anti-Saturn hemisphere and does not stretch far into the trailing one, only covering ~25% of it. So it maybe if our data covered the trailing side too we would see it being darker than the leading one. An albedo change coincident with the thermally anomalous region would not be surprising, since Schenk already observed an increase in Mimas' UV color (0.338 μm) (Schenk et al., 2011). However, we note the opposite trend was observed at shorter wavelengths: Hendrix et al. (2012) observed a decrease in Mimas' albedo in the 0.17 to 0.19 μm wavelength region, which was attributed to the presence of hydrogen peroxide ($H_2O_2$) – a UV darkening agent which can be produced by radiolysis and photolysis of water ice. However, we note that albedo interpretations must be made with caution, as statistically the albedos inside and outside of the anomalous region are indistinguishable: 0.48±0.06 inside and 0.40±0.07 outside. More work is required to validate this result. Within their uncertainty the thermophysical properties we determine are in good agreement with Howett et al. (2011), who found the thermal inertia inside the anomaly to be 66±23 MKS, and outside to be <16 MKS, with albedos across Mimas to range between 0.49 and 0.70.

Figure 7 shows the thermal inertia inside and outside of the anomalous region for different latitude bins. The results cover ~31% of Mimas' surface; the anomaly covers ~24% of Mimas, but this study covers 59% of the anomaly. The results clearly show higher thermal inertias are observed inside of the anomalous region. There also does appear to be a decrease in thermal inertia towards higher latitudes both inside and outside of the anomalous region. However, they also show that at some latitudes (e.g. 0 to 10° N) the errors are so large (particularly on values outside of the anomalous region) that within uncertainties the thermal inertias inside and outside the anomaly are consistent. It is expected that if these error bars were reduced (e.g. by making additional observations at additional local times) this overlap would be lost, and the two values would be distinct with those inside of the anomaly being higher than those outside it.

We find the mean and standard deviation of albedo and thermal inertia across the observed area of Mimas to be 0.43±0.08 and 68±50 MKS respectively. If instead Mimas' leading (trailing) hemispheres are considered separately then the albedo is 0.45±0.08 (0.41±0.07), and the thermal inertia is 98±42 MKS (34±32 MKS). These values are in good agreement with our previously published for Mimas: albedo of $0.49^{+0.05}_{-0.14}$ and thermal inertia of and $19^{+57}_{-9}$ MKS (Howett et al., 2010). This is in part because the uncertainty on Mimas' thermal inertia is so large. We note however that our albedos are lower than those determined by Pitman et al. (2010), who used data from Cassini Visual and Infrared Mapping Spectrometer (VIMS) to determine a global bolometric Bond albedo of Mimas of 0.67±0.10, and leading and trailing hemisphere values of 0.65±0.13 and 0.72±0.10 respectively. We are not sure why such differences occur.

Nordheim et al. (2017) published maps of the electron energy cutoff across Mimas (see Figure 8). Their results confirm that of Paranicas et al. (2014): Mimas' leading hemisphere is preferentially bombarded by high (~MeV) energy electrons. They showed that the *minimum* energy of electrons bombarding the leading hemisphere is between ~1 MeV and >10 MeV, while on the trailing hemisphere the *maximum* energy of the bombarding electrons is ~1 keV to 1 MeV. Nordheim et al. (2017) also published maps of the predicted energy deposition with depth, these results at a depth of 1 cm (i.e. those closest to the skin depth of CIRS) are shown in Figure 9. The results show that while the thermally anomalous region on Mimas' leading hemisphere could form relatively quickly (~6 to 7 million years) one on the leading hemisphere would take >9 million years to form. Nordheim et al. (2017) shows on Mimas' leading hemisphere at shallower depths (<1 cm) this time decreases to <3.3 million years, and increases to >10 millions years deeper in the surface (>2 cm). This short timescale (geologically speaking) may explain why surface alteration by high-energy electrons appears to be the dominant process on Mimas' leading hemisphere.

The predicted differences between the electron bombardment patterns on Mimas' leading and trailing hemisphere are consistent with our results. We detect a thermally anomalous region where it is predicted that the highest energy electrons bombard Mimas' surface, and the total energy accumulation at centimeter depths occurs the fastest (i.e. where the surface is modified the most in any given amount of time). It could be that Mimas' trailing hemisphere will eventually display a similar thermal anomaly when its total energy dose is comparable to the current dose of the leading hemisphere. However, it is also viable that surface processing of Mimas' trailing hemisphere occurs too slowly and that competing surface alteration/weathering processes will prevent the surface here from modifying to the same extent. The most notable competing process to the surface modification by high-energy electrons on Mimas' trailing hemisphere is the deposition of E ring grains, which primarily bombard this hemisphere (c.f. Hamilton and Burns, 1994; Juhász and Horanyi, 2015; Nordheim et al., 2017). These grains will modify and grow the surface too, competing with the surface modification by charged particles (i.e. burying the modified surface).

## 5 Conclusion

We have made new surface temperature maps of Mimas using data taken by Cassini CIRS, which was obtained during four different flybys from 2005 and 2011. All of the maps use only FP3 data (600 to 1100 cm$^{-1}$), since there are no other observations that can provide both high-spatial resolution and are sensitive to the surface temperatures observed. The temperature maps confirm the presence of a thermally anomalous region at low latitudes on Mimas' leading hemisphere, discovered by Howett et al. (2011). The anomaly is warmer during the night and cooler during the day than its surrounding regions.

These surface temperatures are used to derive maps of Mimas' bolometric Bond albedo and thermal inertia by comparing them to modeling results. The thermophysical property maps confirm the anomalous region has a higher thermal inertia than its surroundings (98±42 MKS compared to 34±32 MKS), but its albedo is comparable (0.45±0.08 compared to 0.41±0.07). These results are consistent with the earlier work of Howett et al. (2011), who found the thermal inertia inside (outside) of the anomaly to be 66±23 MKS (<16 MKS) and the albedo to vary between 0.49 and 0.70 across Mimas. The high thermal inertia region is shown to be spatially consistent with a darkening in Mimas' IR/UV color ratio (Schenk et al. 2011) and with where high-energy electrons preferentially bombard Mimas (Nordheim et al., 2017). Thus, these results are consistent with the hypothesis that high-energy electrons are modifying Mimas' surface, changing their UV scattering centers and mobilizing water ice grains (i.e. increasing the conductivity between grains, which in turn increases the surface's thermal inertia).

# 7 Acknowledgements

Thanks are given to the NASA Cassini Data Analysis program, which funded this work (NNX12AC23G and NNN13D466T), to Dr. C. Paranicas for supplying the electron power contours, and to the Cassini Project.

# 8 Tables

| Observation Date (UTC) | Start Time (UTC) | End Time (UTC) | Rev | Spacecraft-Target Distance (km) | Sub-spacecraft Longitude (° W) | Sub-spacecraft Latitude (°) | Sub-Solar Longitude (° W) | Sub-Solar Latitude (°) | Phase (°) at Sub-Spacecraft Location | Mimas-Sun distance (km) |
|---|---|---|---|---|---|---|---|---|---|---|
| 2005/08/02 | 00:59:02 | 01:08:08 | 12 | 113,783 to 109,267 | 179 to 182 | -24 | 221 to 224 | -19 | 45.3 to 44.9 | 1,358,790,092 to 1,358,795,582 |
| 2005/08/02 | 01:51:21 | 02:03:52 | 12 | 90,374 to 85,719 | 196 to 200 | -19 to -17 | 235 to 238 | -19 | 43.1 to 42.7 | 1,358,824,183 to 1,358,833,144 |
| 2005/08/02 | 03:57:45 | 04:07:15 | 12 | 61,831 to 61,390 | 233 to 235 | 12 to 15 | 271 to 274 | -19 | 52.5 to 54.7 | 1,358,922,918 to 1,358,930,654 |
| 2005/08/02 | 05:36:08 | 05:39:27 | 12 | 66,867 to 67,340 | 246 | 42 to 43 | 300 to 301 | -19 | 79.8 to 80.7 | 1,358,999,759 to 1,359,002,143 |
| 2010/02/13 | 19:13:52 | 20:22:12 | 126 | 38,352 to 62,668 | 129 to 156 | -8 to -5 | 147 to 164 | 2 | 22.1 to 12.4 | 1,419,256,058 to 1,419,277,975 |
| 2010/10/16 | 13:56:53 | 14:50:20 | 139 | 118,634 to 96,803 | 159 to 171 | 3 to 5 | 36 to 50 | 5 | 115.2 to 112.6 | 1,430,581,377 to 1,430,614,613 |
| 2011/01/31 | 01:32:10 | 01:49:10 | 144 | 139,099 to 138,496 | 86 to 85 | -2 | 28 to 33 | 7 | 52.3 to 46.9 | 1,435,408,581 to 1,435,416,279 |

Table 1 – Details of the FP3 Cassini/CIRS observations used to make Mimas' albedo and thermal inertia maps.

| Observation Date | Rev | Description |
| --- | --- | --- |
| 050116 | 0C | Small number of FP3 and FP4 north hermisphere stares |
| 050414 & 050415 | 6 | FP3 stare at equatorial region with five detectors |
| 060321 | 22 | FP3 single detector stare at anti-Saturn hemisphere |
| 060630 | 25 | FP3 scan of anti-Saturn hemisphere |
| 060816 | 27 | FP3 stare anti-Saturn hemisphere |
| 061120 | 33 | Very sporadic FP3 observations of southern hemisphere |
| 070527 | 45 | Low spatial resolution FP3 coverage of Saturn-facing southern hemisphere |
| 080323 | 62 | Low spatial resolution sporadic FP3 coverage of Saturn-facing southern hemisphere |
| 080608 | 71 | Low spatial resolution sporadic FP3 coverage of North polar region |
| 080616 | 72 | Low spatial resolution sporadic FP3 coverage of southern hemisphere. |
| 080630 | 74 | FP3 scans of leading southern hemisphere |
| 080714 | 76 | FP3 stare at South pole. |
| **080804** | 79 | FP3 North polar region stares |
| 081024 | 90 | FP3 North polar region stares |
| 090123 | 101 | FP3 North polar region stares low spatial resolution |
| **091014** | 119 | FP3 scan over trailing hemisphere. |
| **100213** | 126 | FP3 scan over anti-sat hemisphere, FP1 anti-Saturn hemisphere coverage |
| **120605** | 167 | High northern hemisphere coverage with FP3 and FP1 |

Table 2 – Other observations CIRS made of Mimas' surface, with a brief description of coverage. Those observations that may be of use to other researchers are highlighted in bold. Specifically these are observations that have dedicated CIRS scans to provide hemispherical coverage at a useable spatial resolution (i.e. comparable to those used in this study).

## 9 Figures

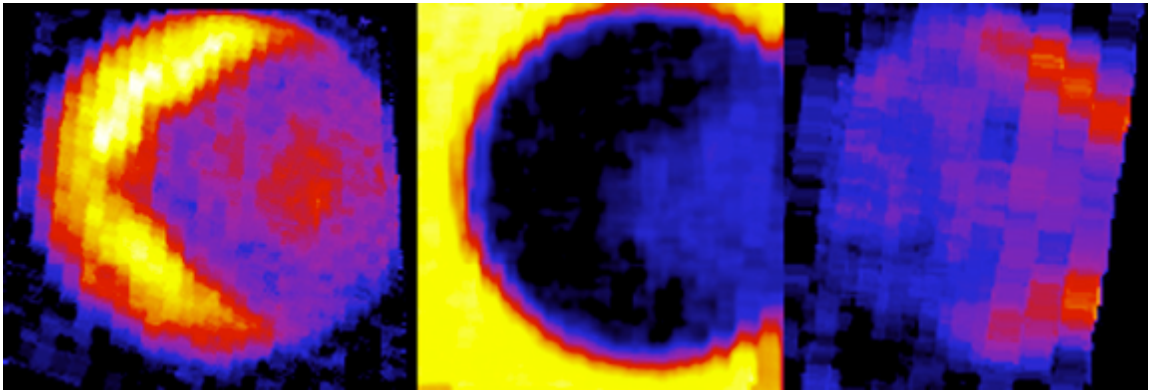

Figure 1 – The "PacMan" anomaly as discovered by Howett et al. (2011). The figure shows 600–650 cm$^{-1}$ CIRS FP3 day and night images of the anti-Saturn side of Mimas, taken in February 2010 (orbit 126) , October 2010 (orbit 139) and January 2011 (orbit 144). The anomalously-cold region appears on the right hand side of the disk in the orbit 126 daytime image (left panel) and on the left side of the disk in the orbit 144 dayside image (right panel), which is centered at more easterly longitudes and thus shows the eastern extent of the anomaly. The same region appears anomalously warm at night (middle panel), indicating a higher thermal inertia than the rest of the disk. Saturn is in the background in the orbit 139 image.

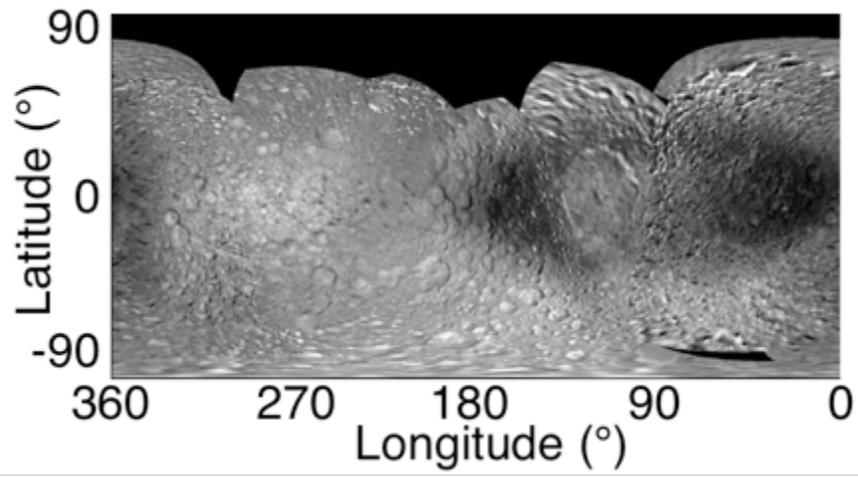

Figure 2 - IR/UV (0.930/0.338 μm) color ratio map of Mimas' surface determine from Cassini ISS data from Schenk et al. (2011).

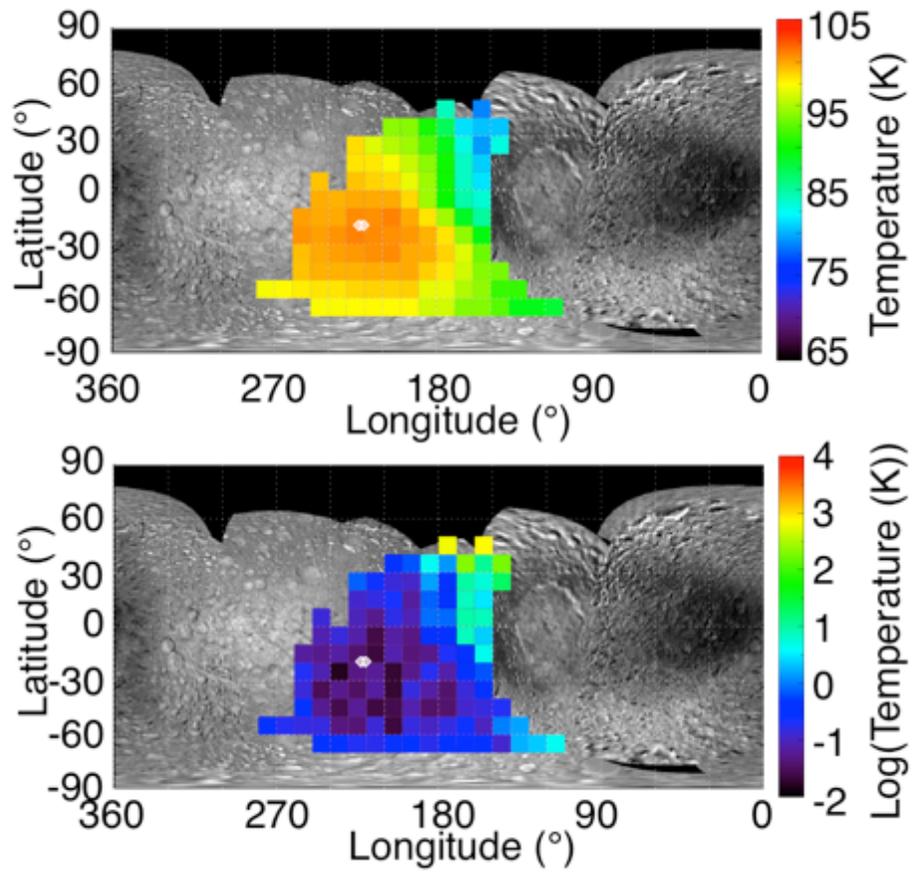

(a) 2$^{nd}$ August 2005 from 00:59:02 to 01:08:08 UTC

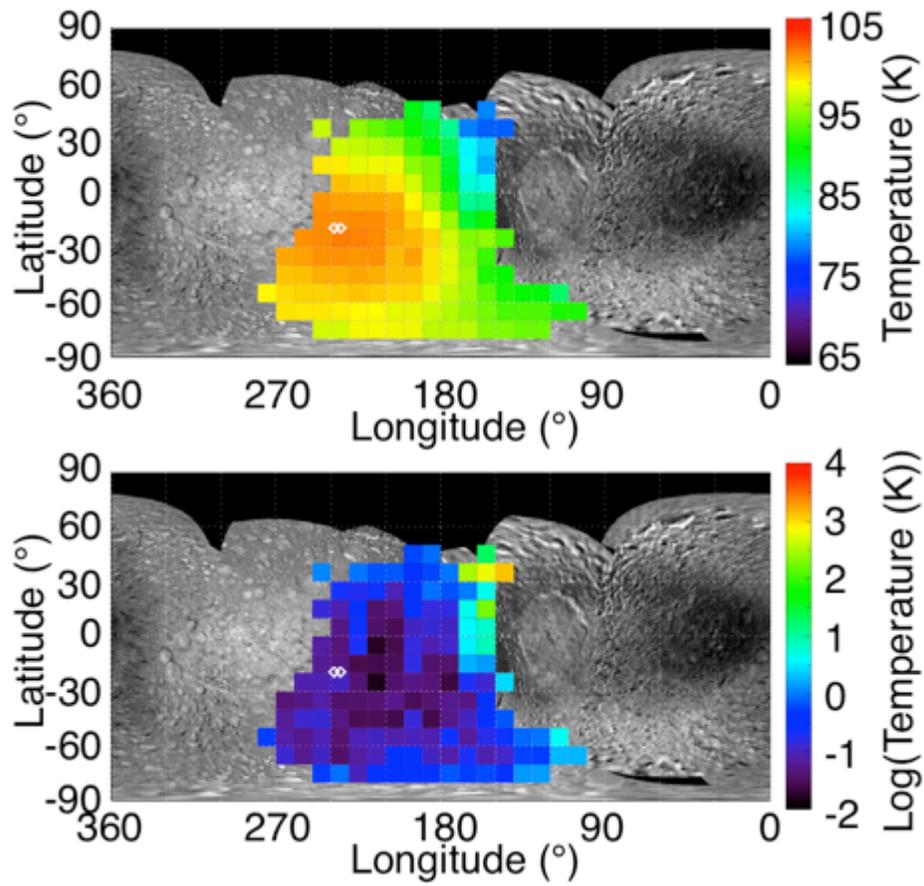

(b) 2$^{nd}$ August 2005 from 01:51:21 to 02:03:52 UTC

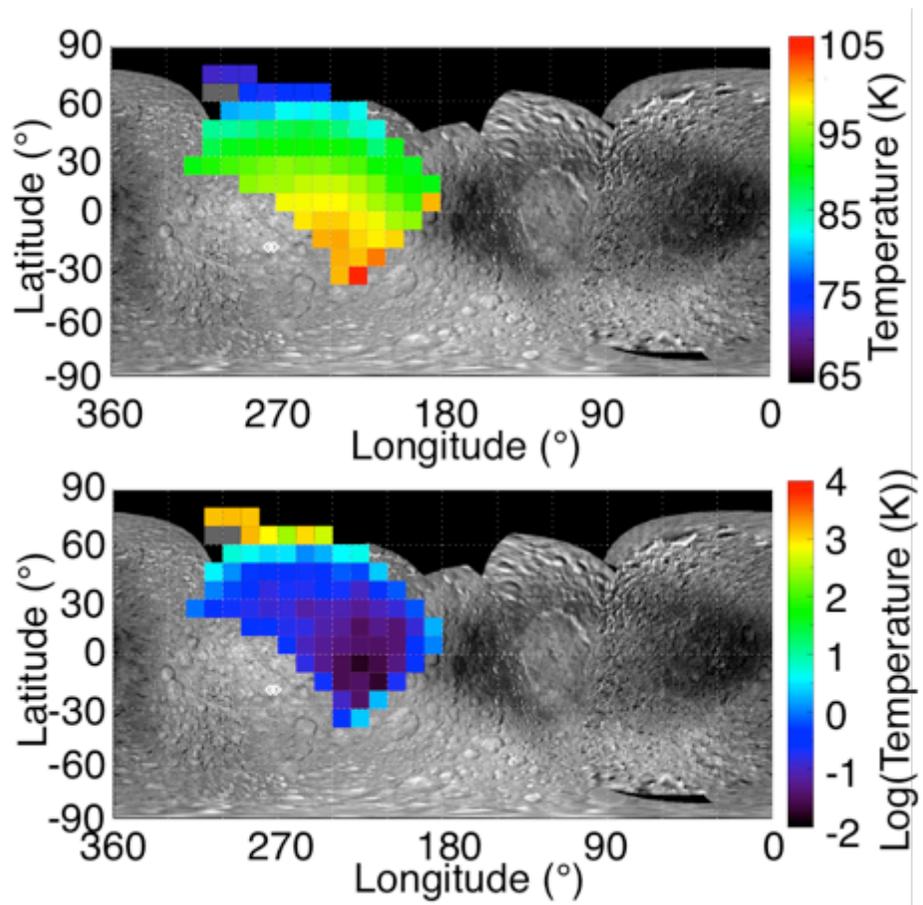

(c) 2nd August 2005 from 03:57:45 to 04:07:15 UTC

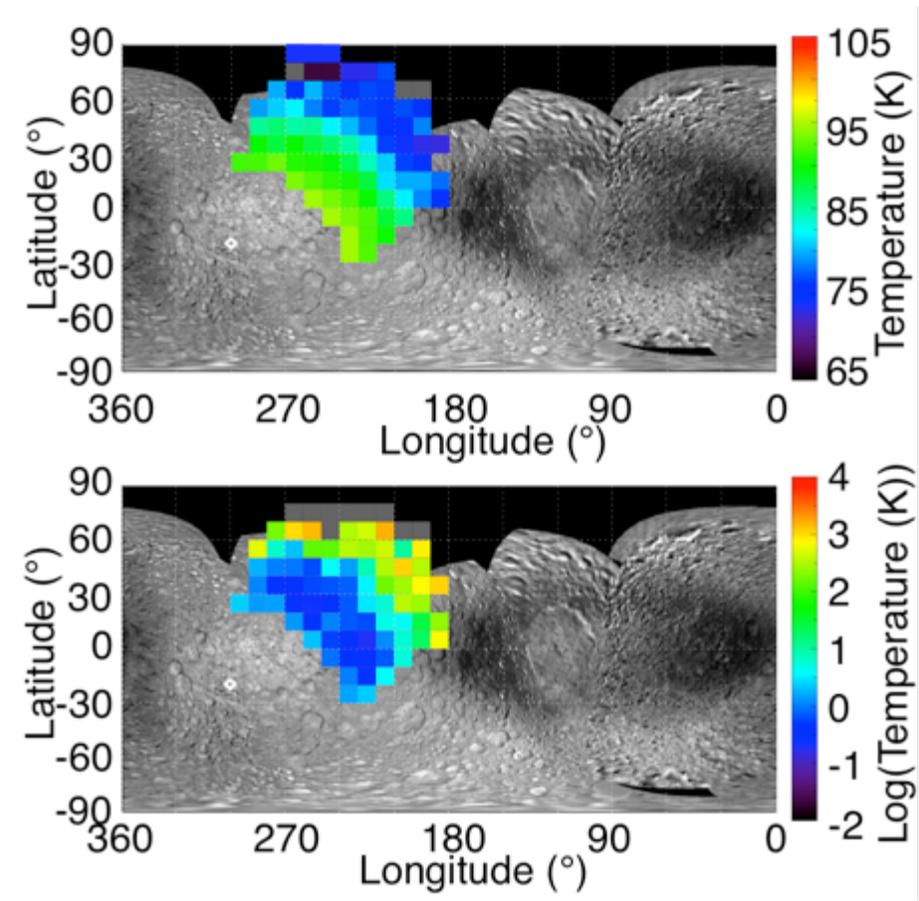

(d) 2nd August 2005 from 05:36:08 to 05:39:27 UTC

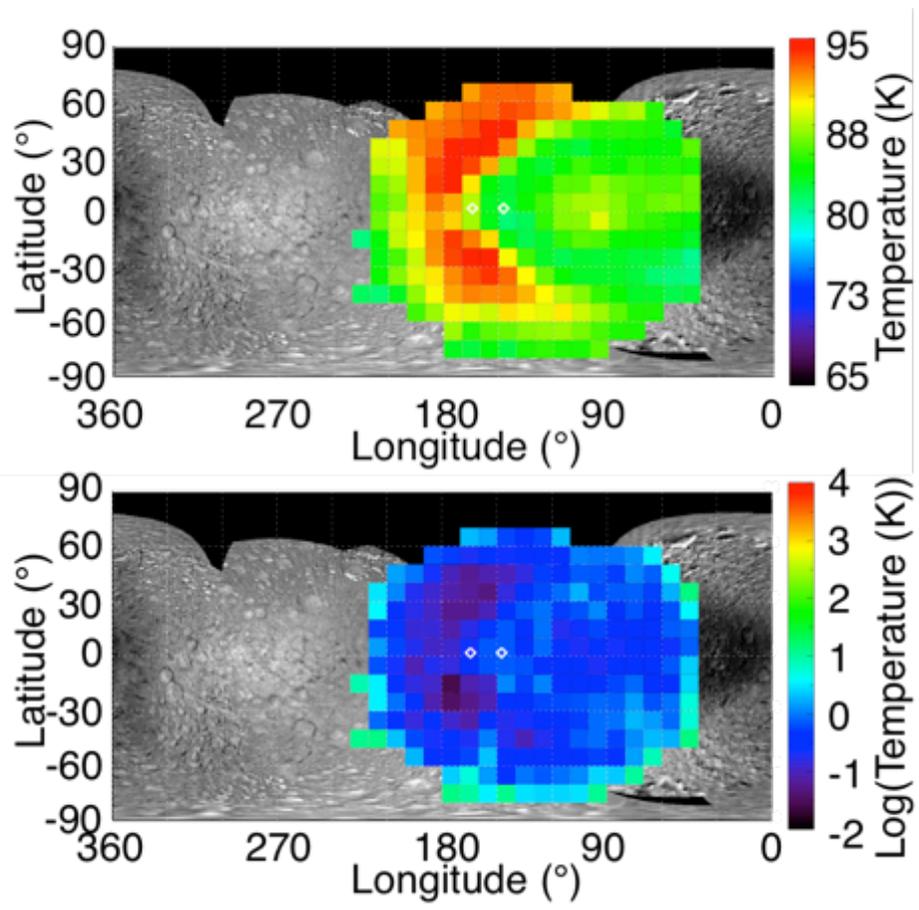

(e) 13th February 2010 19:13:52 to 20:22:12

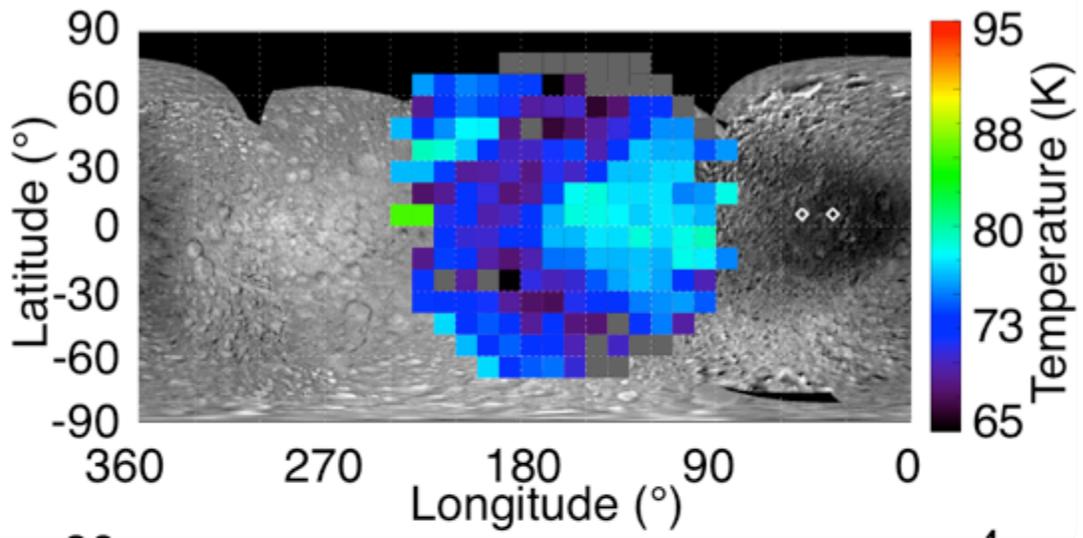

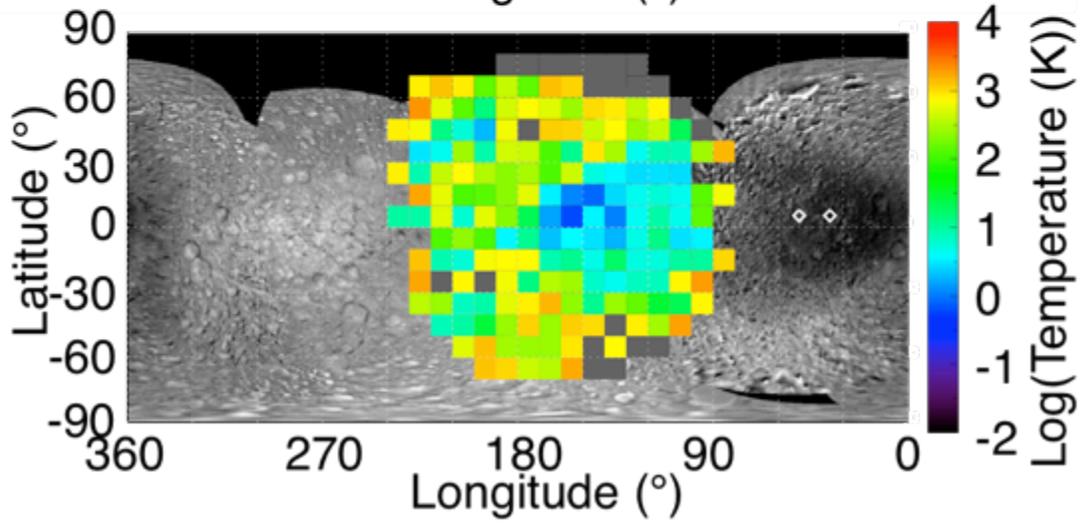

(f) 16th October 2010 13:56:63 to 14:50:20

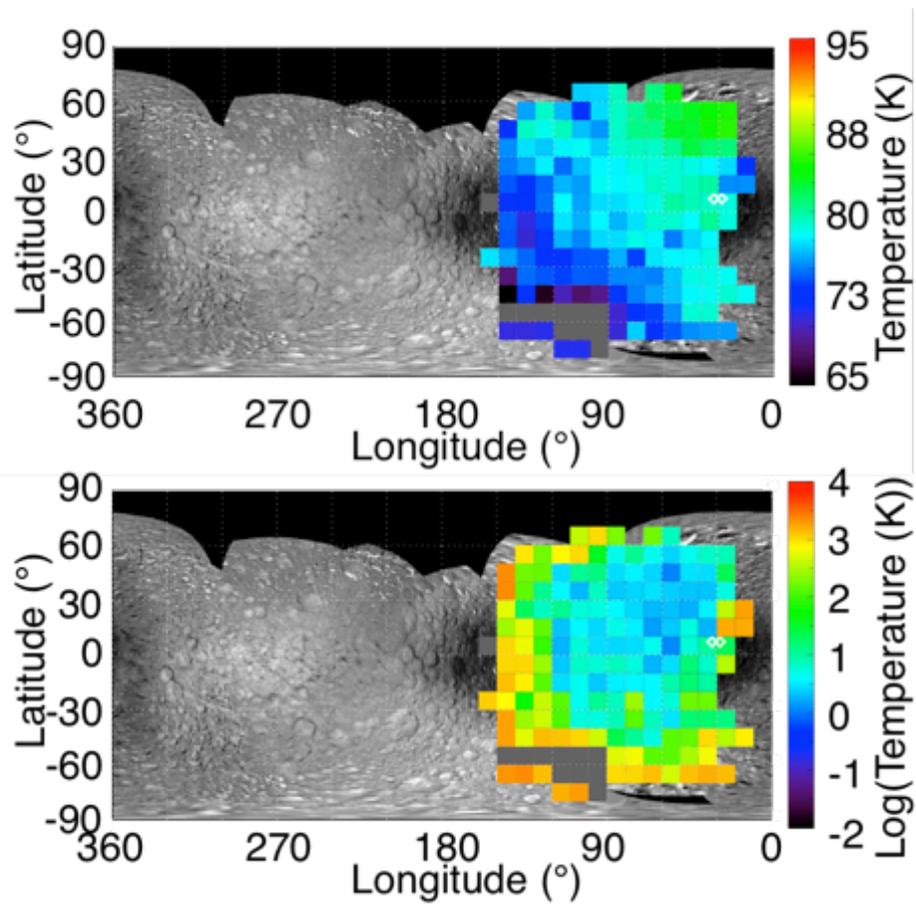

(g) 31st January 2011 01:32:10 to 01:49:10

Figure 3 – Top: Surface temperatures of Mimas derived from Cassini CIRS measurements, taken at different epochs. Bottom: The error of the surface temperatures shown on a logarithmic scale.

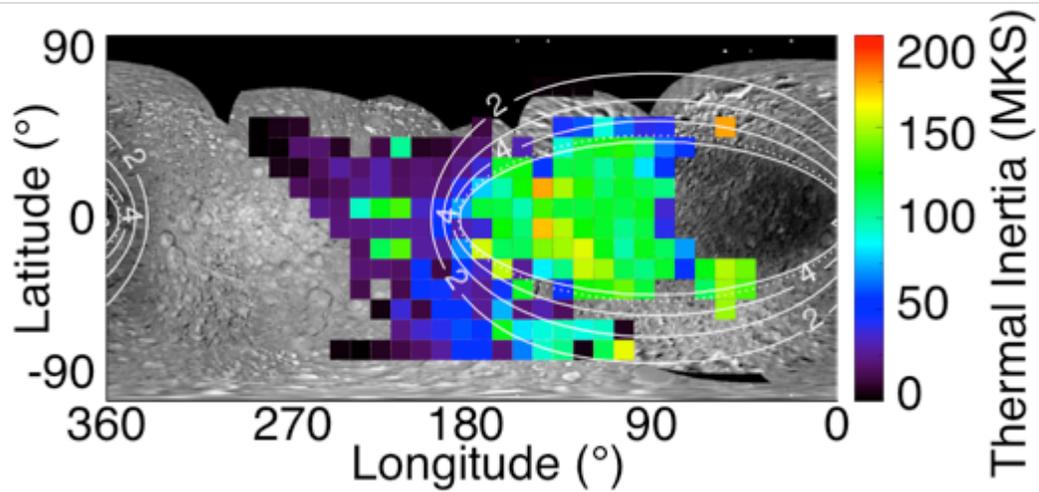

(a) Thermal inertia map of Mimas. Overlaid are contours of energetic electron power deposited into the surface per unit area ($\log_{10}$ MeV cm$^2$ s$^{-1}$) determined using updated results from Cassini's Magnetospheric Imaging Instrument (MIMI).

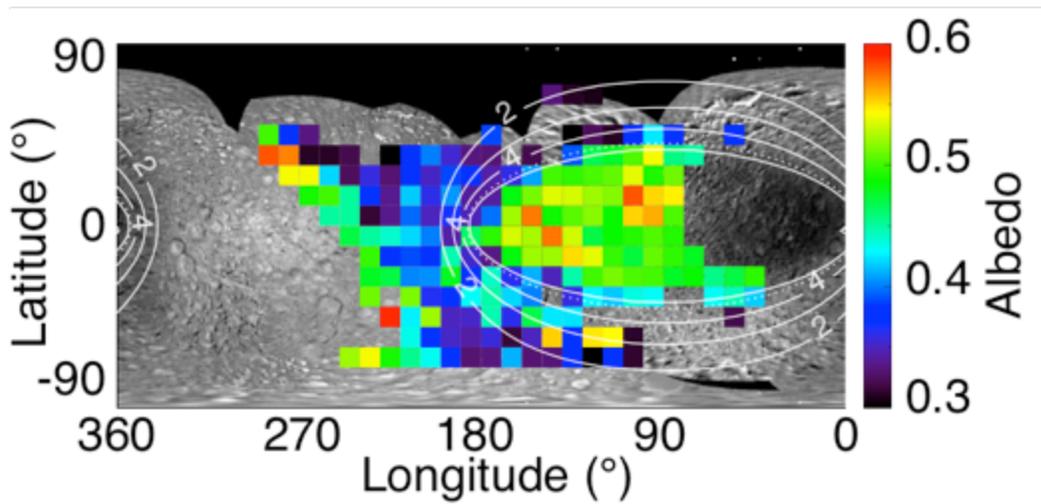

(b) Bolometric Bond Albedo map of Mimas.

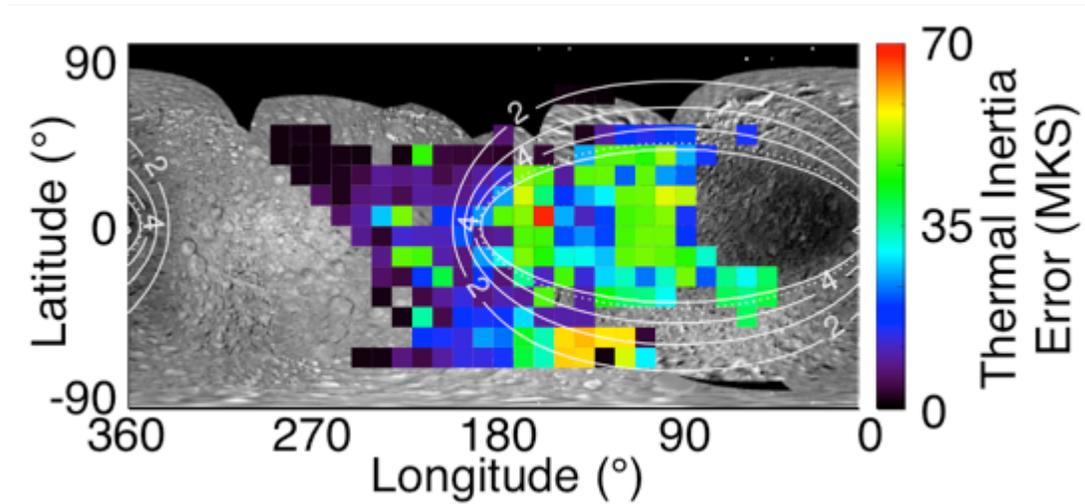

(c) The standard deviation of the thermal inertia values given in (a).

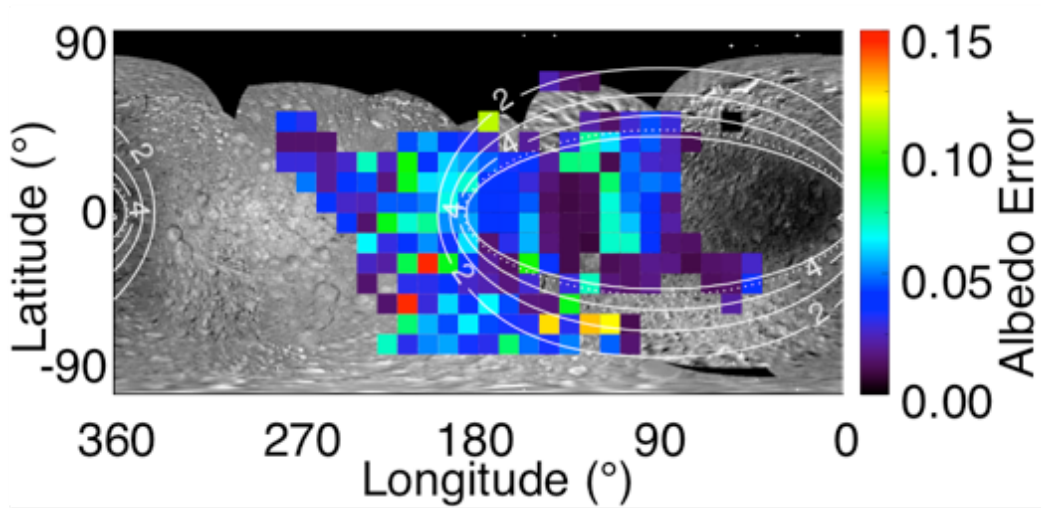

(d) The standard deviation of the albedo values given in (b).

Figure 4 – Maps of derived thermophysical properties for Mimas. The base-map is the IR/UV color map from Schenk et al. (2011). Overlaid are contours of energetic electron power deposited into the surface per unit area (log10 MeV cm$^2$ s$^{-1}$) determined using updated results from Cassini's Magnetospheric Imaging Instrument (MIMI). The best fitting contour to the Mimas color and thermal inertia anomaly boundary (cf. Howett et al., 2011) is given by the dotted line at 4.75 (log10 MeV cm$^2$ s$^{-1}$) or 5.6 × 10$^4$ MeV cm$^2$ s$^-$

1.

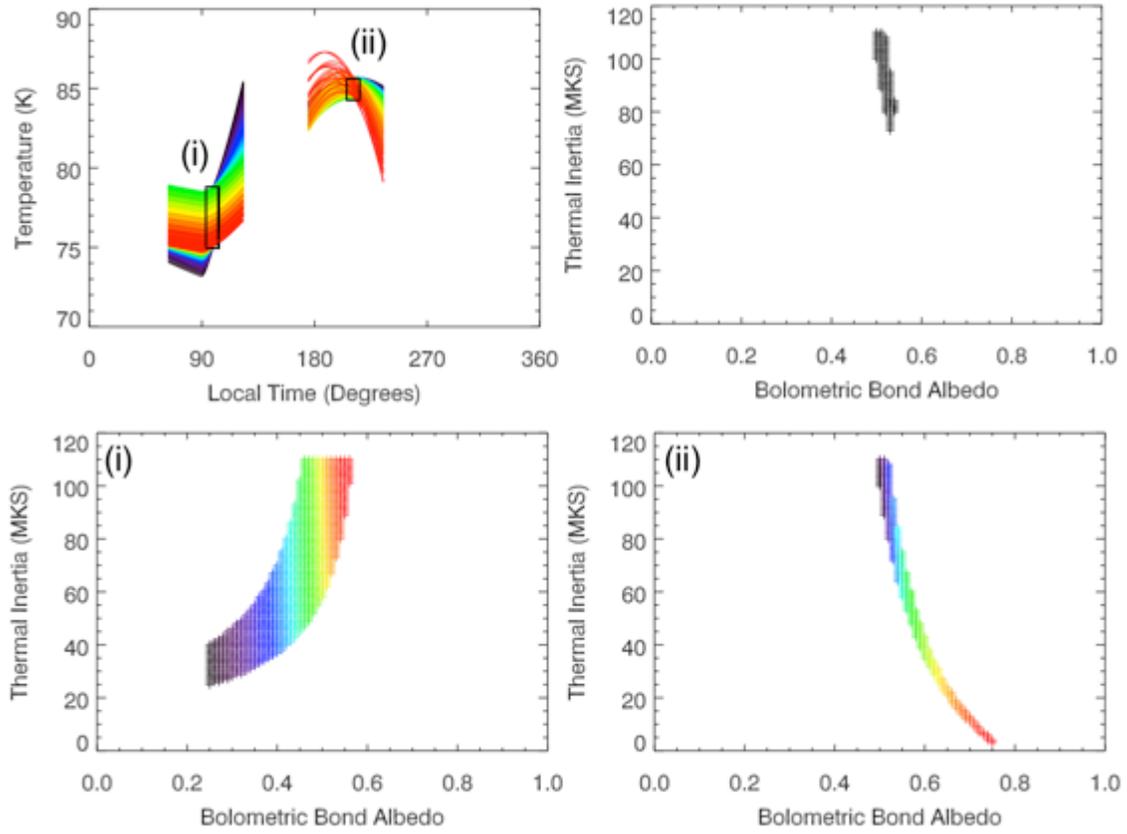

(a) 130° to 140° W, 0° to 10° N

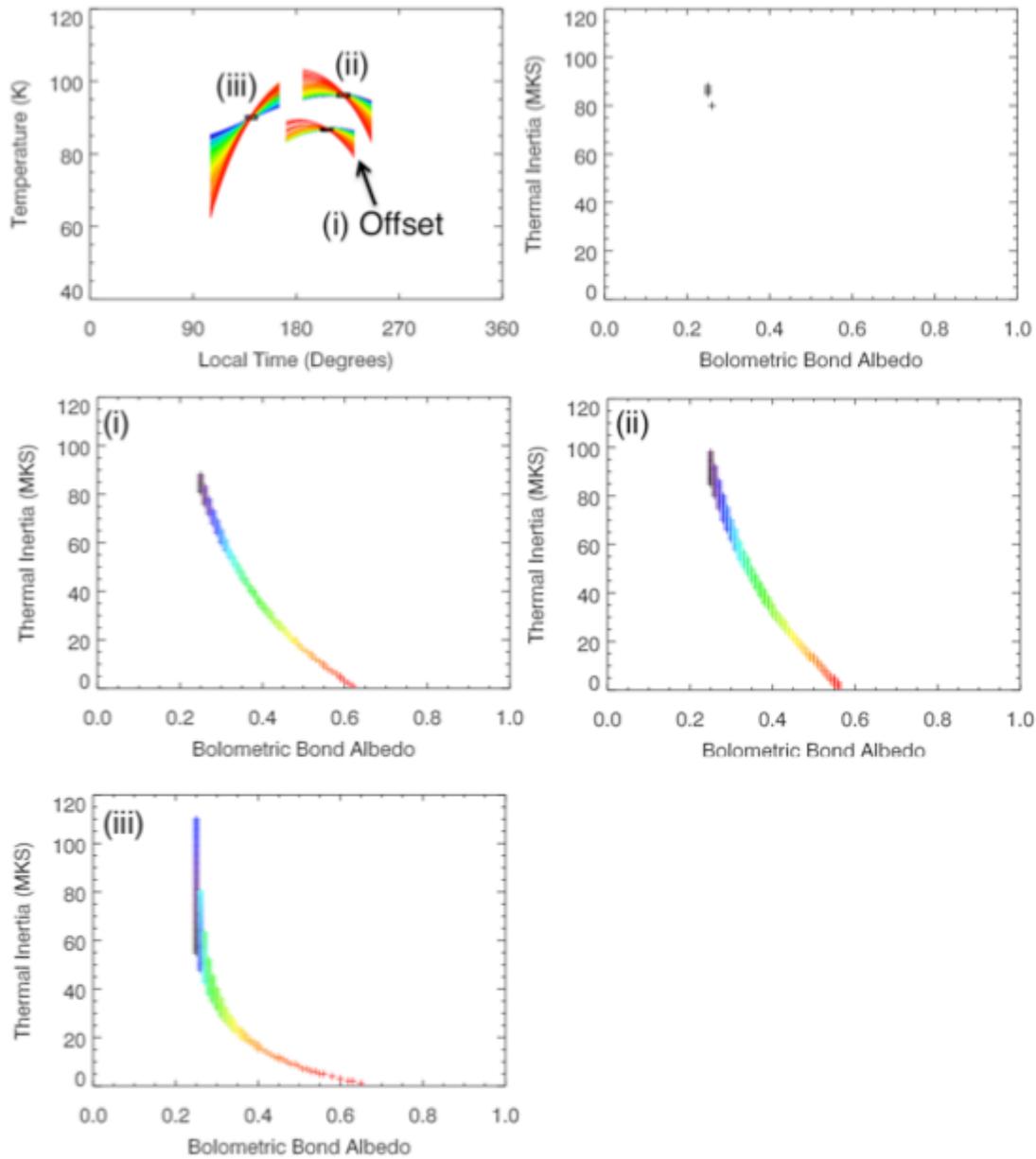

(b) 200° to 210° W, 0° to 10° N

Figure 5 – Details of the model fits to the derived surface temperatures for two locations, one inside the anomalous region (a) and another just outside (b) it. Top Left: The black box shows the surface temperatures observed in a given bin location, accounting for the breadth of local times covered by the bin and the error on the derived temperature. The color chords show the models that are able to fit the observed temperatures, each one

modeled for the geometry of the encounter. Note, in subfigure (b) one of the chords (as indicated by the label and arrow) has been offset to colder temperatures by 10 K for clarity. Top right: The range of thermal inertias and albedos that are able to fit all observations. Other subfigures (i to iii): the thermal inertias and albedos able to fit a single observation, where the color of the point corresponds to the color of the cord shown for that observation (as labeled) in the Top Left subfigure.

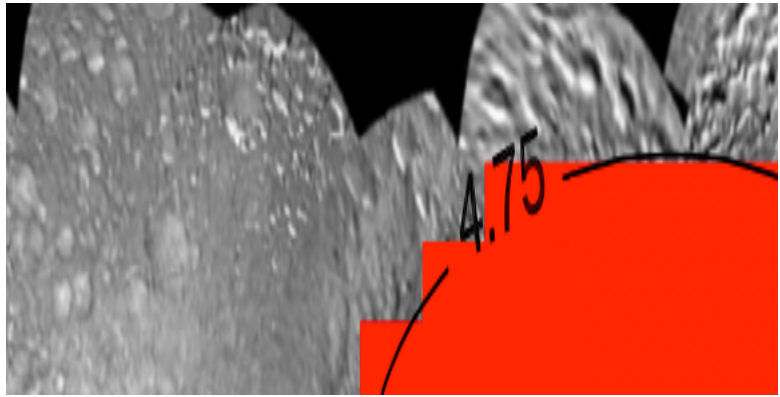

Figure 6 – The bins that are considered "inside" the anomalous region (red), where the boundary is assumed to be the 4.75 log10 MeV cm$^2$ s$^{-1}$ contour energetic electron power deposited ($5.6 \times 10^4$ MeV cm$^2$ s$^{-1}$). "Outside" is assumed to be everywhere that is not colored red.

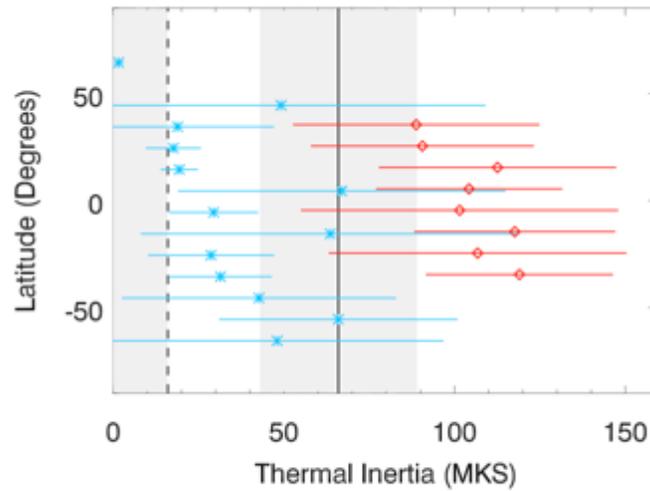

(a) Thermal Inertia variation with latitude. The solid black vertical line gives the previously determined thermal inertia inside of the anomaly with its uncertainty shown by the surrounding grey box; the solid line shows the upper-bound of thermal inertia derived for outside of the anomalous region on Mimas, with the grey region showing the extent of the uncertainty (from Howett et al., 2011)

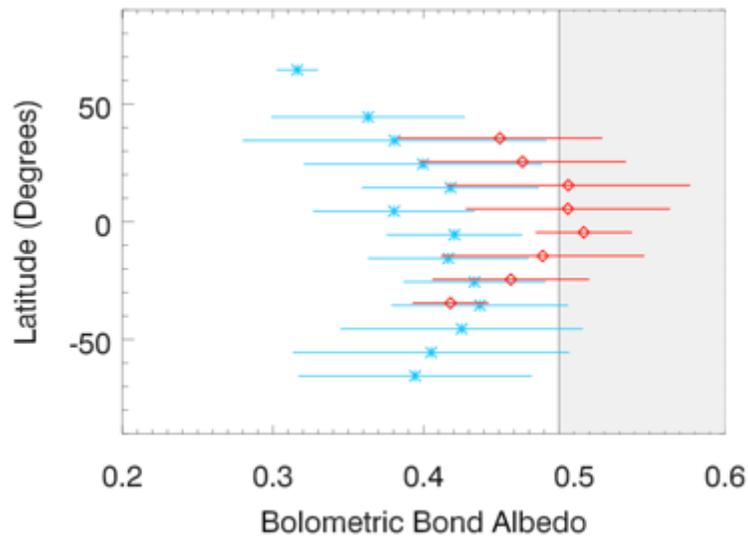

(a) Bolometric Bond albedo variation with latitude. The grey line shows the minimum

bound of the albedos previously observed on Mimas, the range is represented by the grey box but actually extends from 0.49 to 0.70 (i.e. off the scale of this figure) (Howett et al., 2011)

Figure 7 – The mean and standard deviation of thermal inertias and bolometric Bond albedos with latitude. The values are split between those found inside (red, diamonds) and outside (blue, stars) of the anomalous region (see Figure 6 for how inside and outside are defined). Note, the latitudes are shown at the bin center, and the values inside and outside are slightly offset from this for clarity (latitudes outside (inside) the anomaly are decreased (increased) by 0.5°).

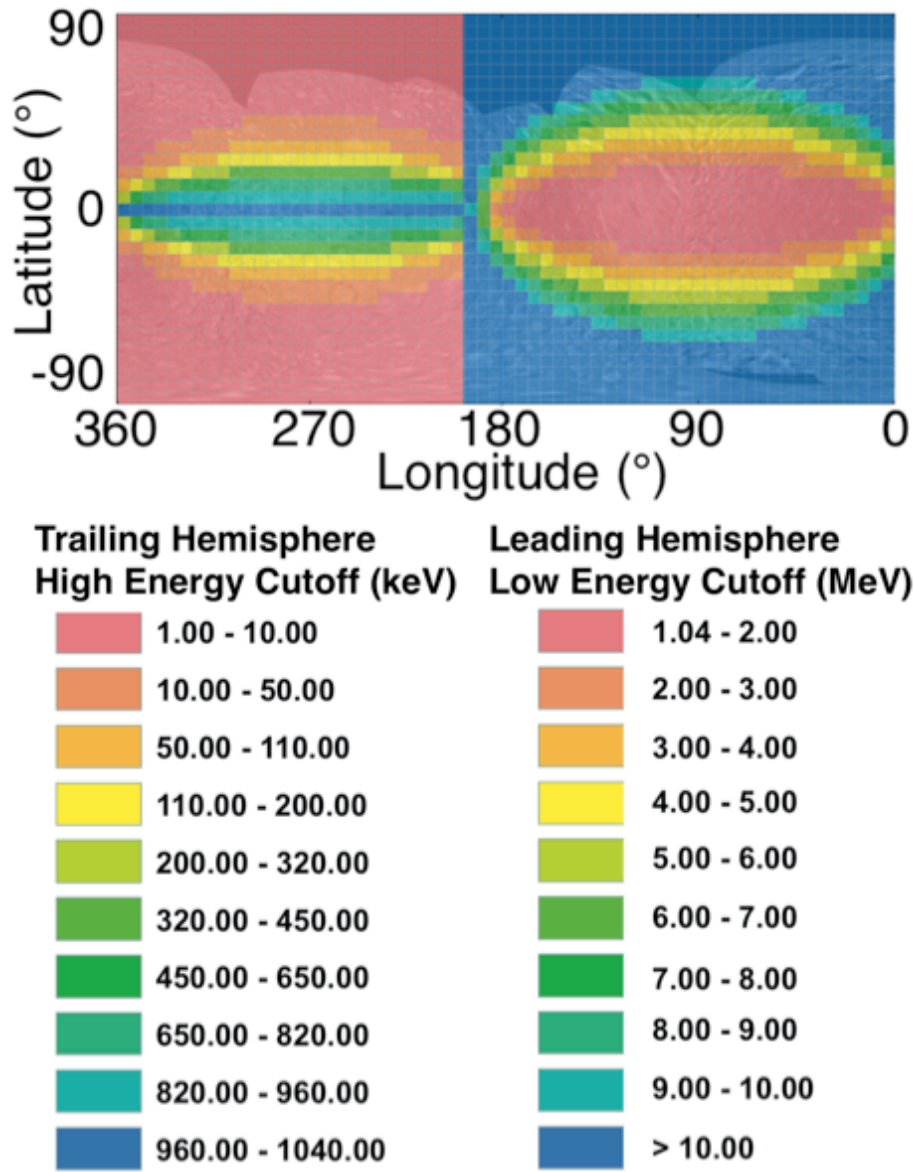

Figure 8: The predicted cutoff energy versus surface location for Mimas, from Nordheim et al. (2017). On the trailing hemisphere the highest energy able to access each location is plotted, whereas on the leading hemisphere the lowest energy is given. The leading hemisphere dose map is overlaid on the trailing hemisphere one. The basemap is from Schenk et al. (2011).

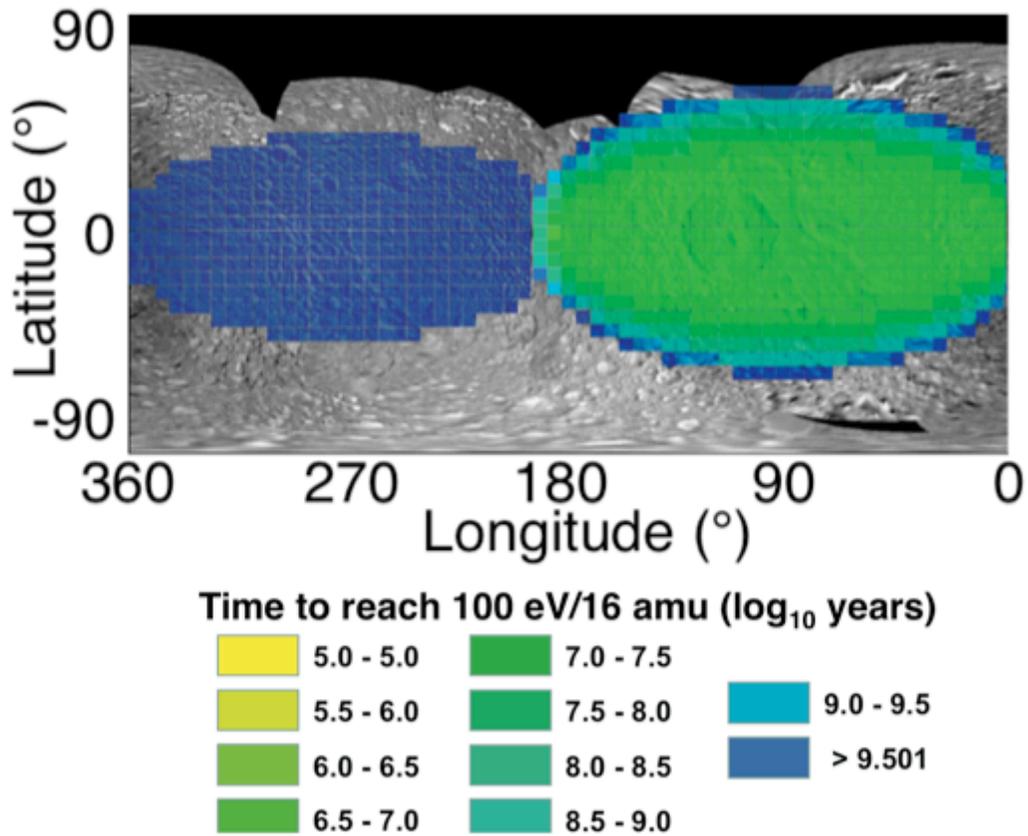

Figure 9: The predicted energy deposition at 1 cm on Mimas (~ the depth probed by CIRS), from Nordheim et al. (2017). The energetic electron dose is given in terms of years to reach a significant dose of 100 eV/ 16 amu, which is equal to a dose of 60.3 Grad. The basemap is from Schenk et al. (2011).